\documentclass[final,5p,times]{elsarticle}

\usepackage{graphicx}	
\usepackage{amsmath}	
\usepackage{amssymb}	
\usepackage{booktabs}
\usepackage{times}
\usepackage{microtype}
\usepackage{epsfig}
\usepackage{caption}
\usepackage{float}
\usepackage{placeins}
\usepackage{color, colortbl}
\usepackage{stfloats}
\usepackage{enumitem}
\usepackage{tabularx}
\usepackage{xstring}
\usepackage{multirow}
\usepackage{xspace}
\usepackage{url}
\usepackage{subcaption}
\usepackage{xcolor}
\usepackage[hang,flushmargin]{footmisc}

\usepackage{algorithm}
\usepackage[noend]{algpseudocode}
\newcommand{\FuncCall}[1]{\texttt{#1}}
\newcommand{\hrulealg}[0]{\vspace{1mm} \hrule \vspace{1mm}}
\usepackage{setspace}

\usepackage{makecell,booktabs,siunitx}

\DeclareMathOperator*{\argmin}{arg\,min}
\newcommand{\norm}[1]{\left\lVert#1\right\rVert}    

\usepackage{lineno}

\begin{document}

\begin{frontmatter}



\title{Snap-and-tune: combining deep learning and test-time optimization for high-fidelity cardiovascular volumetric meshing}

\author[1]{Daniel H. Pak \corref{cor1}}
\cortext[cor1]{Corresponding author}
\ead{daniel.pak@yale.edu}

\author[2]{Shubh Thaker}
\author[2]{Kyle Baylous}
\author[1]{Xiaoran Zhang}
\author[2]{Danny Bluestein}
\author[1]{James S. Duncan}

\affiliation[1]{
    organization={Yale University},
    addressline={300 Cedar St}, 
    city={New Haven},
    postcode={06519}, 
    state={CT},
    country={USA}
}

\affiliation[2]{
    organization={Stony Brook University},
    addressline={100 Nicolls Rd.}, 
    city={Stony Brook},
    postcode={11794}, 
    state={NY},
    country={USA}
}

\begin{abstract}

High-quality volumetric meshing from medical images is a key bottleneck for physics-based simulations in personalized medicine. For volumetric meshing of complex medical structures, recent studies have often utilized deep learning (DL)-based template deformation approaches to enable fast test-time generation with high spatial accuracy. However, these approaches still exhibit limitations, such as limited flexibility at high-curvature areas and unrealistic inter-part distances. In this study, we introduce a simple yet effective snap-and-tune strategy that sequentially applies DL and test-time optimization, which combines fast initial shape fitting with more detailed sample-specific mesh corrections. Our method provides significant improvements in both spatial accuracy and mesh quality, while being fully automated and requiring no additional training labels. Finally, we demonstrate the versatility and usefulness of our newly generated meshes via solid mechanics simulations in two different software platforms. Our code is available at \url{https://github.com/danpak94/Deep-Cardiac-Volumetric-Mesh}.

\end{abstract}


\end{frontmatter}


\section{Introduction}
\label{sec:intro}


Personalized modeling of the human cardiovascular system has been a long-standing interest in medical imaging and biomechanics \cite{taylor2009patient,corral2020digital}. In recent years, purely data-driven approaches have been widely adopted to extract useful information from cardiac imaging \cite{martin2020image,chen2020deep}, but they generally lack interpretability and the ability to predict ``what-if" scenarios in the absence of the appropriate training data. On the other hand, physics-driven approaches can help address these limitations, but they often suffer from the lack of scalable personalization techniques to improve their relevance in real-world clinical scenarios \cite{peirlinck2021precision,xu2021computational}.

A potential bridge for this gap is automated volumetric meshing of patient-specific geometry from medical images \cite{mohamed2004finite,goksel2010image}. Volumetric meshing is the first step to personalization that defines the shape of the object of interest, as well as the computational domain for some of the most well-established simulation techniques in finite element (FE) analyses and fluid-structure interactions (FSI) \cite{bathe2006finite}. However, automated volumetric meshing is notoriously difficult to generalize for high-fidelity medical simulations; in addition to the typical image feature extraction challenges of a segmentation task, meshing also needs to satisfy accurate inter-part connectivity, primitive types and sizes, and many other considerations for successful downstream simulations.

\begin{figure}[!t]
\centerline{\includegraphics[width=\linewidth]{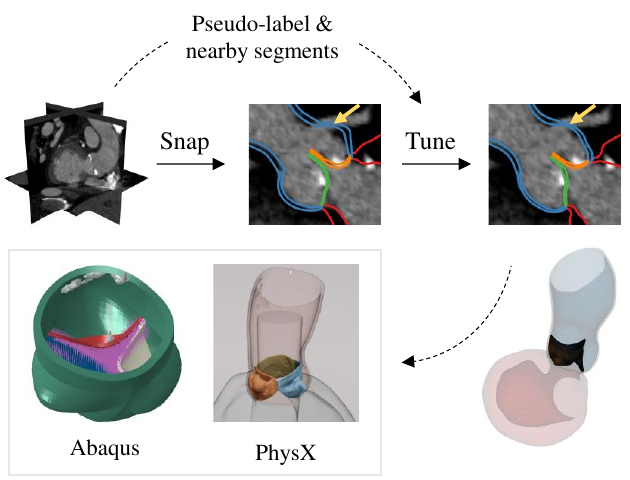}}
\caption{\textbf{Snap-and-tune.} A two-step prediction-and-correction approach for high-fidelity volumetric meshing. Outputs can be directly used in both engineering and graphics simulation software.}
\label{fig:intro_highlight}
\end{figure}

For cardiovascular modeling, this difficulty is further exacerbated by topologically complex structures with low visual cues, such as the calcified vessel wall and heart valves \cite{moccia2018blood,aoyama2022automatic}. These structures are often omitted for simplicity in existing datasets \cite{zhuang2016multi,antonelli2022medical} and segmentation models \cite{isensee2021nnu,wasserthal2023totalsegmentator}, despite their vital role in various clinical scenarios including surgical planning and disease progression \cite{carabello2006aortic,wojnarski2015aortic}. Recently, a series of studies have specifically tackled this problem in cardiovascular modeling by utilizing a combination of deep learning (DL)-based template deformation approaches and mesh attachment techniques \cite{pak2021distortion,pak2023patient,pak2024robust,ozturk2025ai}. In this study, we propose a strategy to further improve these methods.

From qualitative evaluations with multi-center data, we have identified two key limitations in the existing approaches. First, the strain energy-based regularization in training a deformation-based ML model prevents flexible shape adaptation to high-curvature areas, such as the sinuses (i.e. bulges) of the aortic vessel. Second, there is a disconnect between the ML predictions and the mesh attachment step, leading to frequent overestimation of the attached volumes, such as calcium plaques.

To address both of these limitations, we propose a two-step solution, called ``snap-and-tune". [Snap] First, we perform fast shape prediction using a simple forward-pass of a trained DL model. [Tune] Then, we proceed with our test-time optimization step to introduce more detailed sample-specific mesh corrections. Compared to an entirely DL-based approach, our method provides several advantages, such as greater flexibility in the final mesh shape, easier optimization across multiple prediction modules, and better adaptability to test-time user inputs. Compared to an entirely optimization-based approach, our method enables much faster and more stable convergence to a high-quality patient-specific mesh with less user guidance during test-time.

Finally, our method maintains the simulation-readiness of the existing methods and meaningfully affects the final simulation outcome. We demonstrate these points using relevant solid mechanics simulations in engineering (i.e. Abaqus) and graphics (i.e. PhysX) simulation software.


In summary, our contributions are:

\begin{itemize}
    \item Identifying the key limitations of DL-based volumetric meshing for personalized medical simulations
    \item Proposing a novel two-step approach combining deep learning and test-time optimization that addresses these limitations without any additional labels
    \item Demonstrating the versatility and usefulness of the generated meshes using two different simulation platforms
\end{itemize}





\section{Related Works}
\label{sec:related}

Numerous 3D geometry representations have been proposed in the field of vision and graphics, including voxelgrids \cite{schwarz2022voxgraf,sun2022direct}, pointclouds \cite{qi2017pointnet,yu2021pointr}, neural implicit fields \cite{chibane2020neural,molaei2023implicit}, and Gaussian splats \cite{kerbl20233d}. However, these representations are typically not directly utilized in medical simulations due to the lack of high-fidelity simulation techniques and/or infrastructure. Meshless simulation techniques such as smoothed-particle hydrodynamics \cite{monaghan2012smoothed}, physics-informed neural networks \cite{raissi2019physics}, and Simplicits \cite{modi2024simplicits} are exciting areas of research, but their applications in complex multi-component solid and fluid simulations remain to be further developed \cite{yan2016multiphase,sahin2024solving}.

Instead, most representations are first converted into volumetric meshes and plugged into a well-established FE solver \cite{bathe2006finite}, which is often considered the gold standard for large deformation solid mechanics simulations \cite{kim2009skipping,fulton2019latent,modi2024simplicits}. The caveat is that mesh conversions algorithms are nontrivial and error-prone, and are often limited to simple primitives with little to no control over mesh topologies \cite{lorensen1998marching,treece1999regularised,remelli2020meshsdf,chen2004optimal,labelle2007isosurface,hang2015tetgen,hu2020fast}. Options are especially limited for hexahedral meshes, which are the preferred primitive types for many thin-object medical simulations. For complex geometries, meshes often require heavy manual corrections by human experts to be usable for downstream tasks \cite{fabri2009cgal}.

In light of these challenges, studies have tackled the geometry reconstruction problem with a specific focus on high-quality meshing. Numerous approaches have been proposed for processing 2D natural images into 3D meshes with 3D supervision, which places heavy emphasis on the differentiability of the meshing operations and 2D-to-3D feature conversion techniques \cite{liao2018deep,wang2018pixel2mesh,gkioxari2019mesh}. For training without 3D supervision, studies have also frequently employed differentiable rendering techniques \cite{kato2018neural,liu2019soft,tewari2022advances}, which mostly enables high-quality triangular surface meshing \cite{gao2020learning,shen2021deep}. However, additional post-processing is still required for volumetric meshing. Furthermore, neural rendering techniques are difficult to transfer to medical tasks, as the imaging physics and the task characteristics differ significantly.


As such, shape reconstruction for medical simulations often require a unique set of algorithms and insights to achieve the desired level of geometrical accuracy and downstream simulation performance. For cardiovascular modeling, studies have historically relied on heavy assumptions about the anatomy and imaging characteristics to build context-specific meshing algorithms \cite{grbic2013image,labrosse2015subject,ghesu2016marginal}. More recently, these assumptions were relaxed with deep learning approaches \cite{pak2020efficient,kong2022learning,pak2021distortion,pak2023patient}, which uses generalizable feature extraction and template deformation techniques to generate high-quality simulation-ready meshes. A robust mesh attachment algorithm was proposed to enable the incorporation of sporadic objects, such as calcium plaques, to the final patient geometry \cite{pak2024robust}. In this work, we focus on further expanding the capabilities of these recent meshing approaches.
\section{Methods}
\label{sec:methods}

\begin{figure*}[!t]
\centerline{\includegraphics[width=\linewidth]{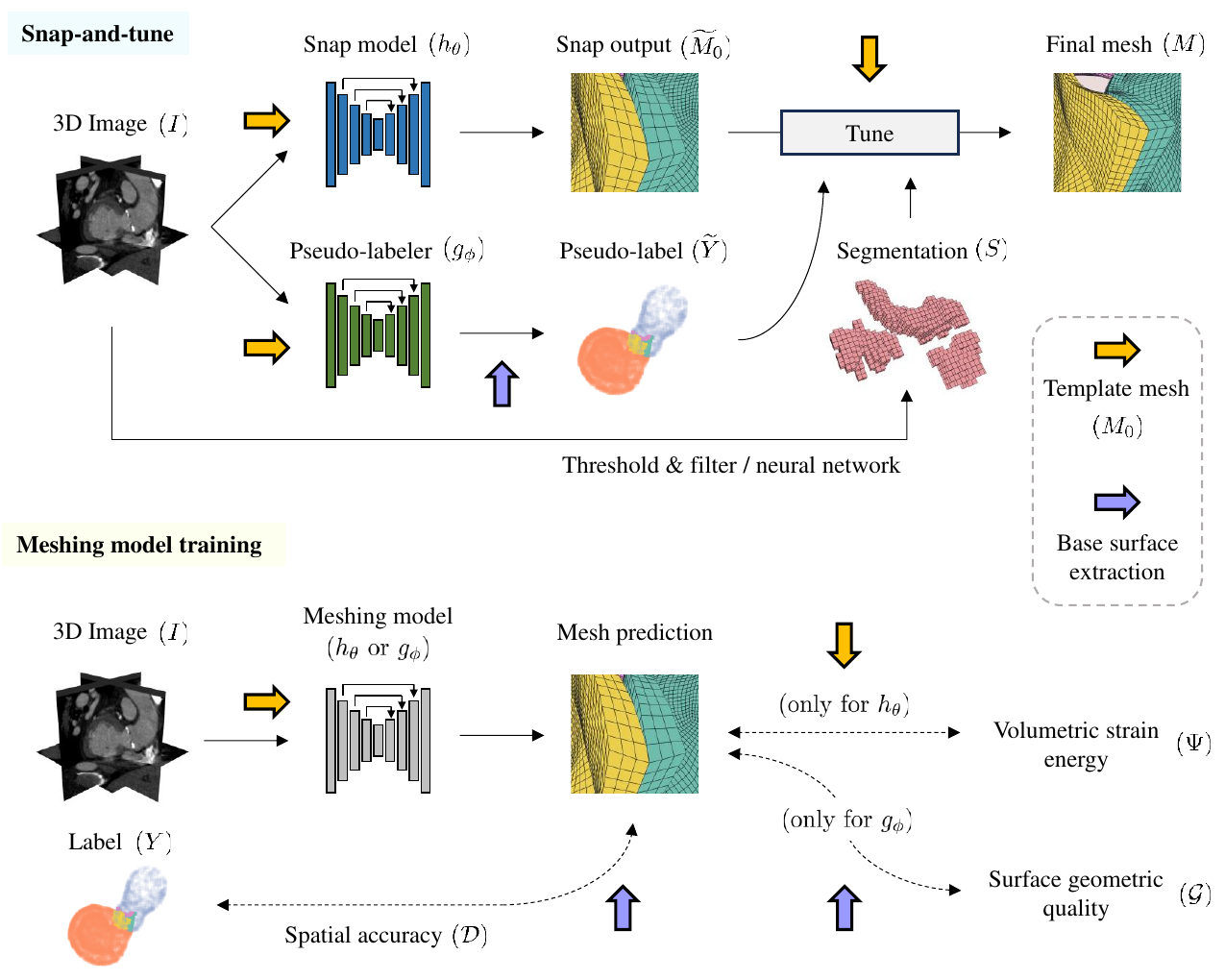}}
\caption{\textbf{Methods overview}. (Top) Snap model predicts an initial volumetric mesh based on the input image. Tune step combines the outputs from two other parallel prediction modules to further improve both the spatial accuracy and element quality of the final mesh. (Bottom) Snap model and the pseudo-labeler are trained using different mesh regularization techniques. Otherwise the training setup is identical.}
\label{fig:methods_main_flowchart}
\end{figure*}

The snap-and-tune method is broadly based on a volumetric mesh deformation strategy (Fig.~\ref{fig:methods_main_flowchart}). Given an initial mesh $M_0 = (V_0, C)$ with vertices $V_0$ and cells $C$, the final mesh $M = (V, C)$ is obtained by translating the vertices with displacements $U$, i.e. $V = V_0 + U$. The general objective is to find the optimal displacements $U^*$ that minimizes both the distance error $\mathcal{D}$ to a target $Y$ and the mesh strain energy $\Psi$ caused by the deformation, such that

\begin{equation} \label{eq:opt_u}
    U^* = \argmin_{U} \mathcal{D}(U, Y; M_0) + \lambda_0 \Psi(U; M_0)
\end{equation}

\noindent where $\lambda_i$ are hyperparameters.

\subsection{Snap: ML-based fast mesh prediction}

For our snap phase, we train a neural network $h_\theta$ that predicts input-dependent displacements with a single forward-pass, i.e. $U = h_\theta(I)$. The optimization is thus performed over the network parameters $\theta$

\begin{equation} \label{eq:opt_nn}
    \theta^* = \argmin_\theta \mathbb{E}_{\Omega} \big[ \mathcal{D}(h_\theta(I), Y; M_0) + \lambda_0 \Psi(h_\theta(I); M_0) \big]
\end{equation}

\noindent where $I$ represents images, and $(I,Y) \sim \Omega$ represents a sample from a training set comprised of image-label pairs.


Following \citep{pak2023patient}, $\mathcal{D}$ is defined as the symmetric chamfer distance between the \textit{base} surface pointcloud of the predicted mesh $X(h_\theta(I); M_0)$ and the corresponding pointcloud labels $Y$. For greater accuracy, the pointclouds are often separated into $N$ separate classes, in which case

\begin{equation} \label{eq:chamfer}
    \mathcal{D} = \frac{1}{2N} \sum_{i=1}^{N} \big[ \mathcal{L}_{sided}(X_i, Y_i) + \mathcal{L}_{sided}(Y_i, X_i) \big]
\end{equation}

\begin{equation}
    \mathcal{L}_{sided}(A, B) = \frac{1}{|A|} \sum_{\mathbf{a} \in A} \min_{\mathbf{b} \in B} \norm{\mathbf{a}-\mathbf{b}}_2^2
\end{equation}

The strain energy is often defined as a function of the deformation gradient $\mathbf{F}$ \cite{kim2020dynamic}. For thin cardiovascular tissues, we use the linear combination of isotropic (as-rigid-as-possible) and anisotropic energy (anisotropic square-root variation of St. Venant Kirchoff) defined as

\begin{equation}
    \Psi = \frac{1}{|C|} \sum_{c \in C} \Big[ \norm{\mathbf{F} - \mathbf{R}}_F^2 + \lambda_1 \Big( \sqrt{\mathbf{d}^T \mathbf{F}^T \mathbf{F} \mathbf{d}} - 1 \Big)^2 \Big]
\end{equation}

\noindent where $\mathbf{R}$ is the rotational component of the polar decomposition of $\mathbf{F}$, $\norm{\cdot}_F$ is the Frobenius norm, and $\mathbf{d}$ is the anisotropy penalty direction. All variables are calculated separately for each cell $c \in C$ ($\mathbf{F}_c$, $\mathbf{R}_c$, $\mathbf{d}_c$), but the subscripts have been omitted for readability. Furthermore, only $\mathbf{F}$ and $\mathbf{R}$ are dependent on $U$; $\mathbf{d}$ is defined as the thickness direction of each element in the initial geometry $M_0$. For more background information, please refer to \cite{kim2020dynamic,pak2023patient}.


\subsection{Tune: test-time optimization for high-fidelity mesh correction}

The tune phase performs direct test-time optimization of displacements

\begin{equation}
    \widetilde{U}^* = \argmin_{\widetilde{U}} \mathcal{\widetilde{D}}(\widetilde{U}, \cdot \; ; \widetilde{M}_0) + \lambda_{user} \widetilde{\Psi}(\widetilde{U}; M_0)
\end{equation}

\noindent where for each subject, we use the mesh generated by $h_\theta$ as our new initial template $\widetilde{M}_0$, and optimize a new $\widetilde{U}$ as the tune-phase displacements. $\lambda_{user}$ is a user-specified weighting hyperparameter during test-time to control the ``stiffness" of the optimized deformation.


The main question then becomes how to define the distance loss and the prediction target during test-time. We propose two sets of loss-target pairs to simultaneously address the rigidity and anatomical inconsistency in the final mesh predictions. Thus, $\widetilde{\mathcal{D}} = \widetilde{\mathcal{D}}_{1} + \lambda_2 \widetilde{\mathcal{D}}_{2}$, where each $\widetilde{\mathcal{D}}_i$ is further described in the following sections.


\textbf{Tune part 1}: We train a second neural network $g_\phi$ that generates pseudo-labels $\widetilde{Y}$, which are used to further refine the snap-phase outputs. Similar to Eq.~\ref{eq:chamfer}, $\widetilde{Y}(g_\phi(I); M_0)$ is the \textit{base} surface pointcloud of the second neural network's predicted mesh. $g_\phi$ is trained using surface geometric regularizers instead of the volumetric strain energy, leading to much greater flexibility and higher spatial accuracy at high-curvature areas

\begin{equation}
    \phi^* = \argmin_\phi \mathbb{E}_{\Omega} \big[ \mathcal{D}(g_\phi(I), Y; M_0) + \mathcal{G}(g_\phi(I); M_0) \big]
\end{equation}

$\mathcal{G} = \lambda_3 \mathcal{L}_{normal} + \lambda_4 \mathcal{L}_{laplacian} + \lambda_5 \mathcal{L}_{edge}$, similar to \cite{wang2018pixel2mesh,pak2021weakly}. $\mathcal{G}$ was only evaluated using the base surface elements. $\mathcal{L}_{normal}$ and $\mathcal{L}_{laplacian}$ are standard implemenations from Pytorch3d, and $\mathcal{L}_{edge}$ is the edge correspondence loss that better utilizes the high-quality template mesh than the standard edge length loss \cite{pak2021weakly}

\begin{equation} \label{eq:edge_correspondence}
\mathcal{L}_{edge} =\; \frac{1}{\mid \varepsilon \mid} \sum_{e_i \in \varepsilon} \Big( \frac{\norm{e_i}_2}{\max\limits_{e_i^\prime \in \varepsilon} \norm{e_i^\prime}_2} - \frac{\norm{U \circ e_i}_2}{\max\limits_{e_i^\prime \in \varepsilon} \norm{U \circ e_i^\prime}_2} \Big)^2
\end{equation}

\noindent Note that we used identical training conditions for $h_\theta$ and $g_\phi$, except for the mesh regularization.

After training, we perform a forward-pass to obtain the pseudo-labels for each test sample $\widetilde{Y}(g_\phi(I); M_0)$, and minimize its discrepancy with the base surface pointcloud of the final optimized mesh $\widetilde{X}(\widetilde{U}; \widetilde{M}_0)$

\begin{equation} \label{eq:tune_D1}
    \widetilde{\mathcal{D}}_{1} = \frac{1}{2N} \sum_{i=1}^{N} \big[ \mathcal{L}_{sided}(\widetilde{X}_i, \widetilde{Y}_i) + \mathcal{L}_{sided}(\widetilde{Y}_i, \widetilde{X}_i) \big]
\end{equation}

\noindent This loss-target pair serves to introduce greater flexibility and higher spatial accuracy of the final volumetric meshes.

\textbf{Tune part 2}: Next, we aim to address the inter-part inconsistency of the snap-phase outputs. For cardiovascular modeling, this is especially important for minimizing the error in the reconstructed calcium plaque volume \cite{pak2024robust}, but our approach is broadly applicable to minimizing undesired inter-part distances between any neighboring objects.

Let us first assume we have obtained the voxelgrid segmentation of the neighboring object. For calcification, this can be done using simple thresholding and filtering operations as well as deep learning models. Then, we convert the segmentation into surface meshes $S$ using standard isosurface extraction techniques. From the surface meshes, we select a subset of surface elements for optimization using a series of mesh-based operations (Alg.~\ref{alg:tune_2}).

\begin{algorithm}
    \caption{Grouping and filtering surface elements} \label{alg:tune_2}
    \begin{algorithmic}[1]
    \Function{GroupAndFilterElems}{$\widetilde{M}_0, S$}
    \State $\{(\widetilde{M}_i, S_i)\}_{i=1}^{\hat{N}} \gets \FuncCall{AssignPairs}(\widetilde{M}_0, S)$
    \For {$(\widetilde{M}_i, S_i) \textrm{ in } \{(\widetilde{M}_i, S_i)\}_{i=1}^{\hat{N}}$} 
        \State $\mathbf{D} \gets \FuncCall{NearestPointDirection}(\widetilde{M}_i, S_i)$
        \State $\mathbf{N} \gets \FuncCall{SurfaceNormal}(\widetilde{M}_i)$
        \State $S_i \gets \FuncCall{FilterByDirection}(S_i, \mathbf{D}, \mathbf{N})$
        \State $S_i \gets \FuncCall{FilterByDistance}(S_i, \mathbf{D})$
    \EndFor
    \State \Return{$\{(\widetilde{M}_i, S_i)\}_{i=1}^{\hat{N}}$}
    \EndFunction
    \end{algorithmic}
    \hrulealg
    $S$: surface mesh of the neighboring object(s)
\end{algorithm}

Alg.~\ref{alg:tune_2} is the pseudo-code for the mesh-based operations. \FuncCall{AssignPairs} pairs each of the $\hat{N}$ connected components of $S$ to a mesh component in $\widetilde{M}_0$ with the lowest chamfer distance. From each pair of $(\widetilde{M}_i, S_i)$, $\FuncCall{NearestPointDirection}$ finds the difference vector of each point from $\widetilde{M}_i$ to  the nearest neighboring point in $S_i$. $\FuncCall{SurfaceNormal}$ outputs the point-wise surface normal at each node of $\widetilde{M}_i$. $\FuncCall{FilterByDirection}$ discards points on $S_i$ that do not satisfy a cosine similarity threshold between the nearest point direction and surface normal. $\FuncCall{FilterByDistance}$ discards $S_i$ points that are too far from $\widetilde{M}_i$. When the points are discarded, the associated elements of $S_i$ are also discarded.

After the grouping and filtering operations, we define $\hat{X}_i(\widetilde{U}; \widetilde{M}_0)$ as the surface pointcloud of the \textit{entire} surface of the $i^{\textrm{th}}$ mesh component, and $\hat{Y}_i$ as the surface pointcloud of each processed $S_i$. Then, the inter-part distance is minimized using the one-sided surface distance loss

\begin{equation} \label{eq:tune_D2}
    \widetilde{D}_2 = \frac{1}{\hat{N}} \sum_{i=1}^{\hat{N}} \mathcal{L}_{sided}(\hat{Y}_i, \hat{X}_i)
\end{equation}

\noindent which only minimizes the nearest neighbor distance from the attachment object (i.e. calcification) to the optimized volumetric mesh outputs. This loss-target pair significantly reduces the undesired inter-part distances between neighboring objects that should be in contact.

To summarize the tune phase, $\mathcal{D}$ in Eq.~\ref{eq:opt_u} is replaced with $\widetilde{\mathcal{D}} = \lambda_1 \widetilde{\mathcal{D}}_{1} + \lambda_2 \widetilde{\mathcal{D}}_{2}$, the template mesh is re-intialized as the snap-phase output $\widetilde{M}_0$, and a new $\widetilde{U}$ is directly optimized to obtain the final volumetric mesh $M$.
\section{Results}
\label{sec:results}

\subsection{Data acquisition and preprocessing}


We used the same dataset and preprocessing techniques as previous works \cite{pak2023patient,pak2024robust}. Briefly, the dataset consists of 79 cardiac computed tomography angiography (CTA) scans, 14 from the MM-WHS public dataset and 65 from the private dataset from Hartford Hospital. For the Hartford patients, ethics approvals were granted by the Institutional Review Board (IRB) under Assurance No. FWA00021932 and IRB-Panel\_A under IRB No. HHC-2018-0042. We used 35/9/35 scans for training/validation/testing. Human-annotated labels were available in the form of surface meshes for cardiovascular tissues and voxelgrid segmentation for calcification. The template mesh $M_0$ consists of the left ventricular myocardium, ascending aorta, and aortic valve leaflets, and it was generated with representative anatomical parameters using Solidworks and Hypermesh. Please refer to \cite{pak2023patient,pak2024robust} for further details on the dataset characteristics and labeling procedures.

For preprocessing, we first cropped all images and labels to a spatial resolution of 128$^3$ voxels with 1.25mm$^3$ isotropic spacing, and then applied an elastic deformation augmentation. The cropping was performed with the crop center at the center of each subject's label bounding box, with random 3D Gaussian noise added for translation augmentation. For training $h_\theta$, the image intensities were first clipped to [-158,864] Hounsefield Units (HU), and further min-max normalized to [0,1].

\subsection{Implementation details}

The neural network $h_\theta$ was a modified 3D U-net, as described in \cite{pak2021distortion,pak2023patient}.  $h_\theta$ was trained for 8k epochs, and $g_\phi$ was fine-tuned from $h_\theta$ without any changes in the architecture for 4k additional epochs. Notably, the network outputs were control point displacements within a regularly spaced sparse grid spanning the image space, and the final dense displacements $U$ were obtained using b-spline interpolation and the scaling-and-squaring algorithm to enforce smoothness and diffeomorphism. Tune-phase displacements $\widetilde{U}$ were obtained using the same process, meaning our test-time optimization was ultimately performed on the control point displacements. We used the Adam optimizer for all optimization, with a learning rate of 1e-4 for the neural networks and 1e-3 for the test-time optimization. Hyperparameters $\{\lambda_i\}_{i=0}^5 = \{1, 10, 1, 10, 1, 0.01\}$, which were determined via grid search. $\lambda_{user}$ is adjustable by the end user during test-time, but the default suggested value is 1.

\subsection{Baseline experiments}

The baseline experiments are summarized in Table~\ref{table:exps}. With our experiments, we aimed to evaluate the strengths of snap-and-tune vs. snap-only and tune-only strategies, for all combinations of volumetric mesh regularization vs. a standard field smoothness regularization techniques.

The ``snap only: vol" condition is considered the state-of-the-art (SOTA) for our task, given its previous success in outperforming all other baseline DL approaches involving various label manipulation and optimization techniques \cite{pak2023patient}. The ``tune-only" and ``S \& T" methods are all newly implemented baselines for more thorough comparisons against our final proposed method (``S \& T: vv).

In Table~\ref{table:exps}, the ``smooth" conditions were implemented using the same displacement optimization procedure as the ``vol mesh" conditions, but with the regularization swapped from volumetric mesh strain energy to displacement field bending energy \cite{rueckert1999nonrigid,pak2021weakly}. For the tune-only experiments, we used $g_{\phi}$ to predict the pseudo-labels similar to the snap-and-tune experiments, and ``pre-aligned" the template to match them. The pre-alignment was an affine transform followed by an initial deformation optimization. The initial deformation was optimized with the identical setup as the final tune step, but with a coarser control point spacing to enable more stable optimization with no degenerate elements.

\begin{table}[t!]
\centering

\robustify\bfseries
\def\sym#1{\ifmmode^{#1}\else\(^{#1}\)\fi}

\sisetup{
    detect-weight,
    mode=text,
    separate-uncertainty = true,
    table-align-text-post = false,
    table-align-uncertainty = false,
}

\begin{tabular*}{\linewidth}{
    l @{\extracolsep{\fill}}
    c
    c
}

\toprule

{\makecell{}} & {\makecell{Snap}} & {\makecell{Tune}} \\

\midrule



Snap only: smooth \cite{dalca2019unsupervised} & smooth & N/A  \\

\makecell[l]{Snap only: vol \cite{pak2023patient} \\ (SOTA)} & vol mesh & N/A   \\

\midrule[0pt]

Tune only: smooth & N/A & pre-align \& smooth  \\

Tune only: vol & N/A & pre-align \& vol mesh \\

\midrule[0pt]

S \& T: ss & smooth & smooth \\

S \& T: sv & smooth & vol mesh \\

S \& T: vs & vol mesh & smooth \\

S \& T: vv (ours) & vol mesh & vol mesh \\

\midrule

\end{tabular*}

\caption{\textbf{Outline of experimental conditions.} S \& T: abbreviation for snap-and-tune.}

\vspace{-0.05in}

\label{table:exps}

\end{table}

\begin{figure}[!t]
\centerline{\includegraphics[width=\linewidth]{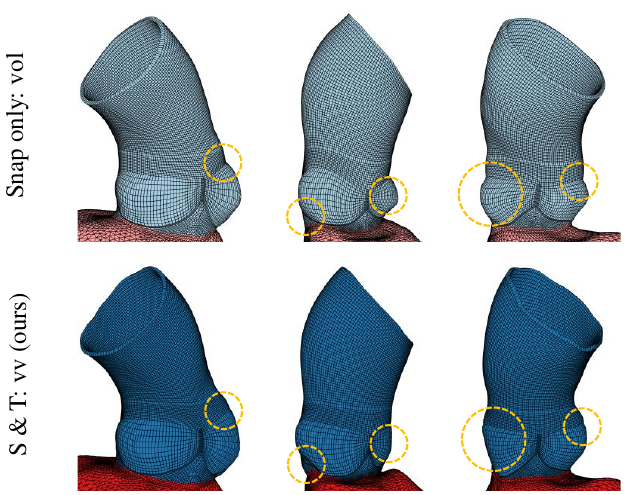}}
\caption{\textbf{Greater flexibility.} Snap-only predictions tend to be slightly rigid and maintain the overall shape of the template mesh, especially around the high-curvature areas. Snap-and-tune provides much more flexibility in those regions while further improving the mesh accuracy and element quality.}
\label{fig:results_heart_3d}
\end{figure}

\subsection{Volumetric mesh predictions} \label{sec:vol_mesh_pred}

\begin{figure}[!t]
\centerline{\includegraphics[width=\linewidth]{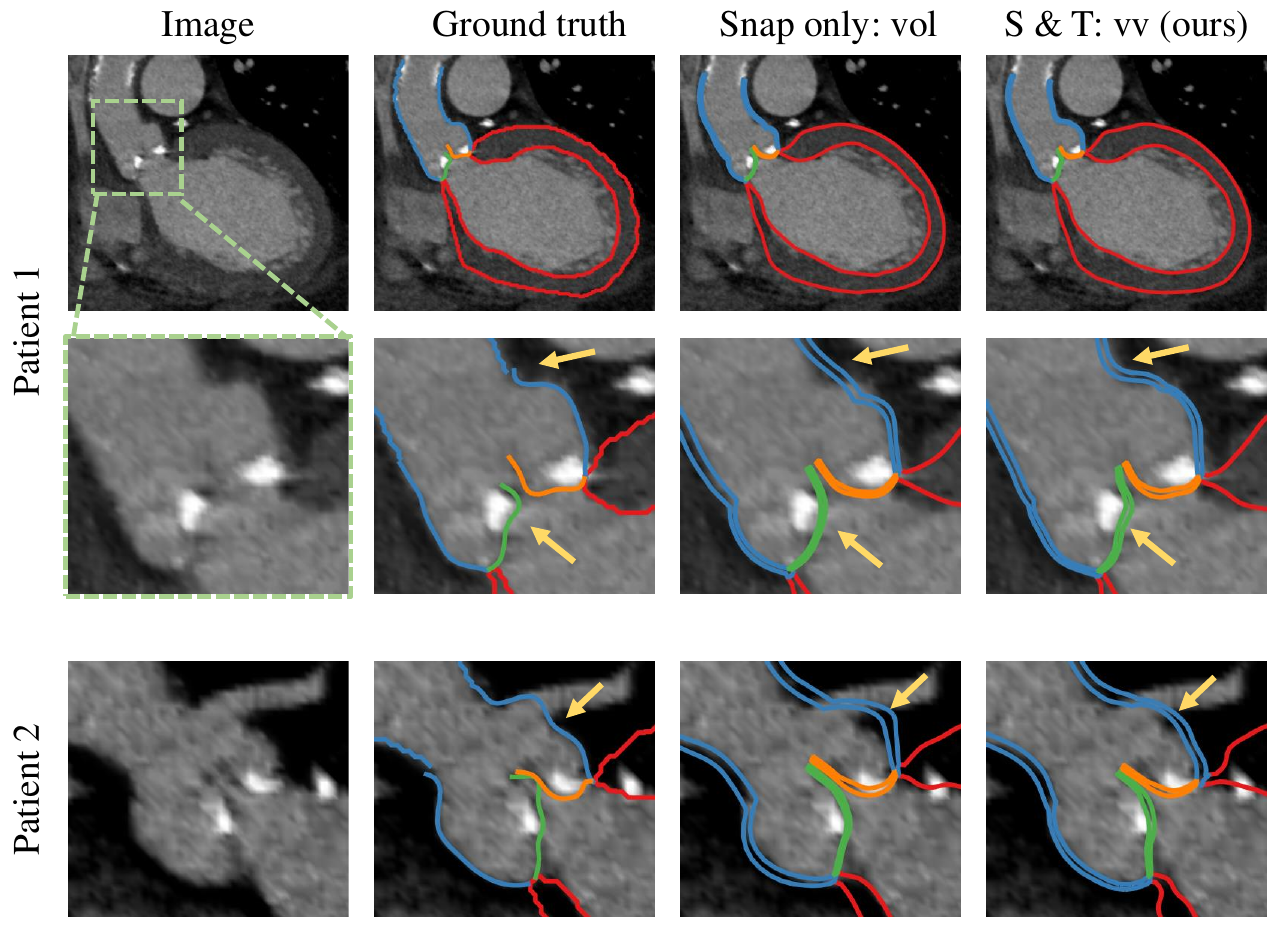}}
\caption{\textbf{Improved volumetric mesh spatial accuracy.} Snap-and-tune improves the spatial accuracy of the generated meshes in high-curvature areas, e.g. bulges in the blood vessel (blue).}
\label{fig:results_img_mesh_slices}
\end{figure}

\begin{figure}[!t]
\centerline{\includegraphics[width=\linewidth]{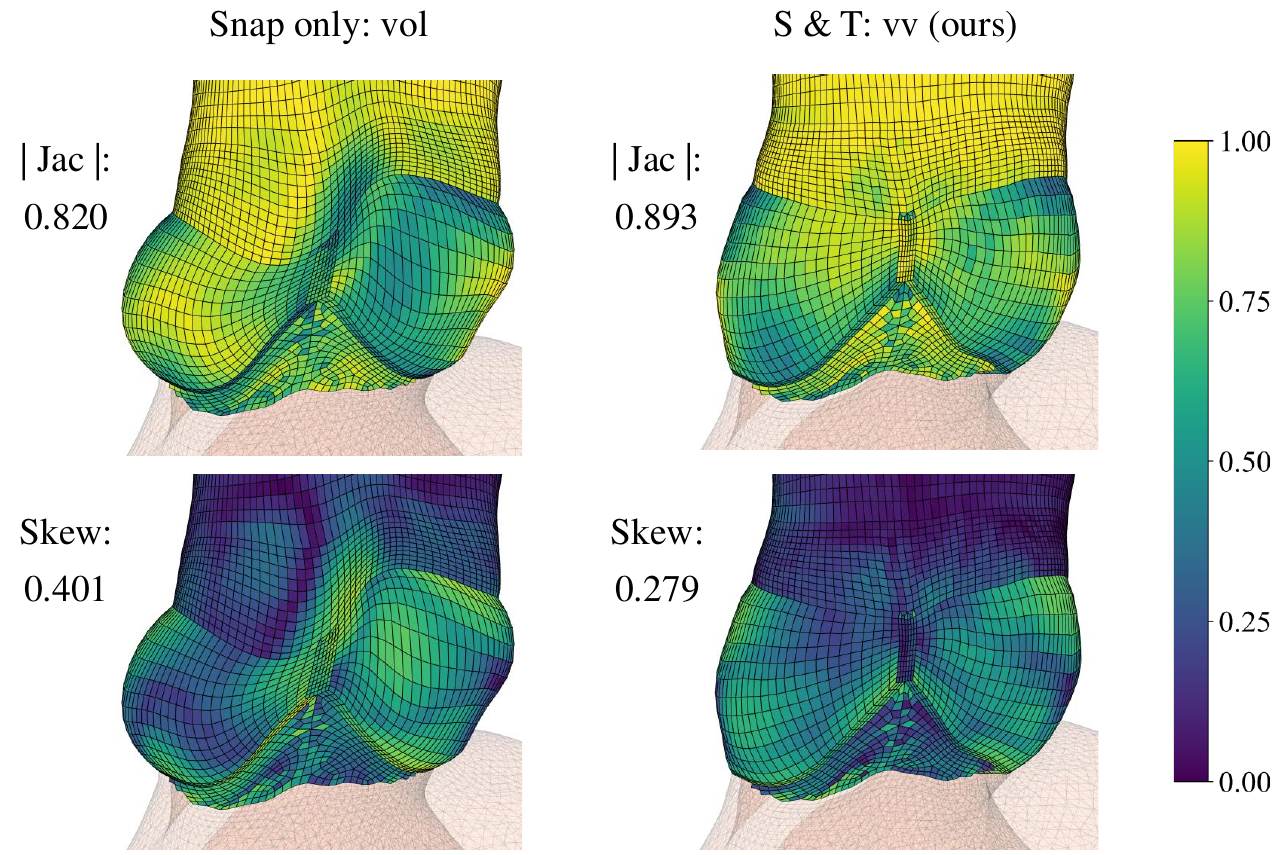}}
\caption{\textbf{Element distortion corrections.} Snap-and-tune can easily recover from large erroneous deformations of snap-only predictions. Top heatmap: scaled jacobian determinant, bottom heatmap: skew. Numbers indicate mean values across all elements.}
\label{fig:results_heart_elem_qual}
\end{figure}

We evaluated both the spacial accuracy and element quality of each algorithm's volumetric mesh predictions, similar to \cite{pak2021distortion,pak2023patient}. As shown in Table~\ref{table:heart}, our final proposed method (``S \& T: vv") significantly outperforms the previous state-of-the-art method (``Snap only: vol") in all metrics by a large margin. The improved accuracy and element quality can also be qualitatively observed in Fig.~\ref{fig:results_img_mesh_slices} and Fig.~\ref{fig:results_heart_elem_qual}, respectively. The performance boost can be partially attributed to the greater flexibility in the deformed meshes (Fig.~\ref{fig:results_heart_3d}), which can heavily affect accurate alignment along high-curvature areas, e.g. sinuses of the vessel wall. As shown in Fig.~\ref{fig:results_heart_3d} (top row), the snap-only DL predictions tend to undesirably maintain the overall curvature of the template mesh, presumably to optimize the mean strain energy metric across the highly variable shapes in the deformation-augmented training set.

The performance improvements are more nuanced between our newly implemented baselines, which were designed specifically to stress test the strengths of our final proposed method. First, volumetric mesh regularization during the tune phase is a clear winner over field smoothness regularization, evidenced by relative improvements in all metrics (Table~\ref{table:heart} odd rows vs. even rows). The only potential competitor to our final proposed method is ``Tune only: vol", but this method takes approximately 2.5x longer run-time while providing less spatially accurate \& higher element quality predictions. Given the importance of the meshes' spatial accuracy in high-fidelity simulations, we argue that our final method is better overall in terms of both speed and meshing performance.

\begin{table*}[t!]
\centering

\robustify\bfseries
\def\sym#1{\ifmmode^{#1}\else\(^{#1}\)\fi}

\sisetup{
    detect-weight,
    mode=text,
    separate-uncertainty = true,
    table-align-text-post = false,
    table-align-uncertainty = false,
}

\begin{tabular*}{\linewidth}{
    @{\extracolsep{\fill}}
    l
    S[table-format=1.3(3)]
    S[table-format=1.3(3)]
    S[table-format=1.3(3)]
    S[table-format=2.2(3)]
    S[table-format=2.3(3)]
    c
}

\toprule

{\makecell{Method}} & {\makecell{CD (mm) $\downarrow$}} & {\makecell{HD (mm) $\downarrow$}} & {\makecell{Thick err (mm) $\downarrow$}} & {\makecell{$|\textrm{Jac}|$ $\uparrow$}} & {\makecell{Skew $\downarrow$}} & {RT (sec) $\downarrow$} \\

\midrule

Snap only: smooth & 2.105 \pm 0.500 & 6.099 \pm 3.847 & 0.146 \pm 0.042 & 0.866 \pm 0.025 & 0.358 \pm 0.039 & 0.3 \\
Snap only: vol    & 1.034 \pm 0.133 & 4.510 \pm 1.767 & 0.051 \pm 0.008 & 0.852 \pm 0.021 & 0.356 \pm 0.027 & 0.3 \\

\midrule[0pt]

Tune only: smooth & 1.008 \pm 0.115 & 4.485 \pm 1.708 & 0.236 \pm 0.081 & 0.830 \pm 0.033 & 0.415 \pm 0.047 & 77 \\
Tune only: vol    & \color{blue} 0.996 \pm 0.136 & \color{red} 4.424 \pm 1.737 & \bfseries 0.015 \pm 0.002 $^{\dagger}$ & \bfseries 0.895 \pm 0.016 $^{\dagger}$ & \bfseries 0.284 \pm 0.023 $^{\dagger}$ & 95 \\

\midrule[0pt]

S \& T: ss        & 1.416 \pm 0.286 & 5.168 \pm 2.806 & 0.158 \pm 0.045 & 0.856 \pm 0.026 & 0.378 \pm 0.039 & 23 \\
S \& T: sv        & 1.036 \pm 0.164 & 4.596 \pm 1.971 & \color{red} 0.018 \pm 0.003 & \color{blue} 0.887 \pm 0.018 $^{\ddagger}$ & \color{red} 0.306 \pm 0.028 & 35 \\
S \& T: vs        & \color{red} 1.001 \pm 0.121 & \bfseries 4.394 \pm 1.749 & 0.053 \pm 0.009 & 0.848 \pm 0.022 & 0.362 \pm 0.029 & 22 \\
S \& T: vv (ours) & \bfseries 0.989 \pm 0.122 $^{*\ddagger}$ & \color{blue} 4.410 \pm 1.690 $^{\ddagger}$ & \color{blue} 0.016 \pm 0.001 $^{*\ddagger}$ & \color{red} 0.883 \pm 0.020 $^*$ & \color{blue} 0.299 \pm 0.029 $^{*\ddagger}$ & 38 \\

\midrule

\end{tabular*}

\caption{\textbf{Spatial accuracy and mesh quality metrics for the predicted volumetric meshes} (i.e. cardiovascular tissues). CD: chamfer distance, HD: Hausdorff distance, Thick err: thickness error, $|\textrm{Jac}|$: scaled Jacobian determinant, RT: run-time. Bold, blue, red: 1$^{\textrm{st}}$, 2$^{\textrm{nd}}$, 3$^{\textrm{rd}}$ best. $p < 0.05$ from ``S \& T: vv (ours)" to ``Snap only: vol" ($*$), ``Tune only: vol" ($\dagger$), and ``S \& T: sv" ($\ddagger$).}

\vspace{-0.05in}

\label{table:heart}

\end{table*}

\subsection{Mesh attachment}

Mesh attachment is a highly useful technique for increasing the accuracy of downstream simulations, especially for a common medical scenario such as calcium plaque formation. We performed the mesh attachment with Calcium Meshing with Anatomical Consistency (C-MAC) \cite{pak2024robust}, which essentially converts a voxel segmentation into a tetrahedral mesh with exact conforming nodal attachments to an existing volumetric mesh \cite{pak2024robust}. 

We evaluated the spatial accuracy and inter-part consistency of the C-MAC outputs following each volumetric meshing algorithm, similar to \cite{pak2024robust}. The number of merged nodes is used as an inter-part consistency metric because calcium plaques should form on top of a tissue (i.e. our predicted volumetric meshes).

\begin{figure}[!t]
\centerline{\includegraphics[width=\linewidth]{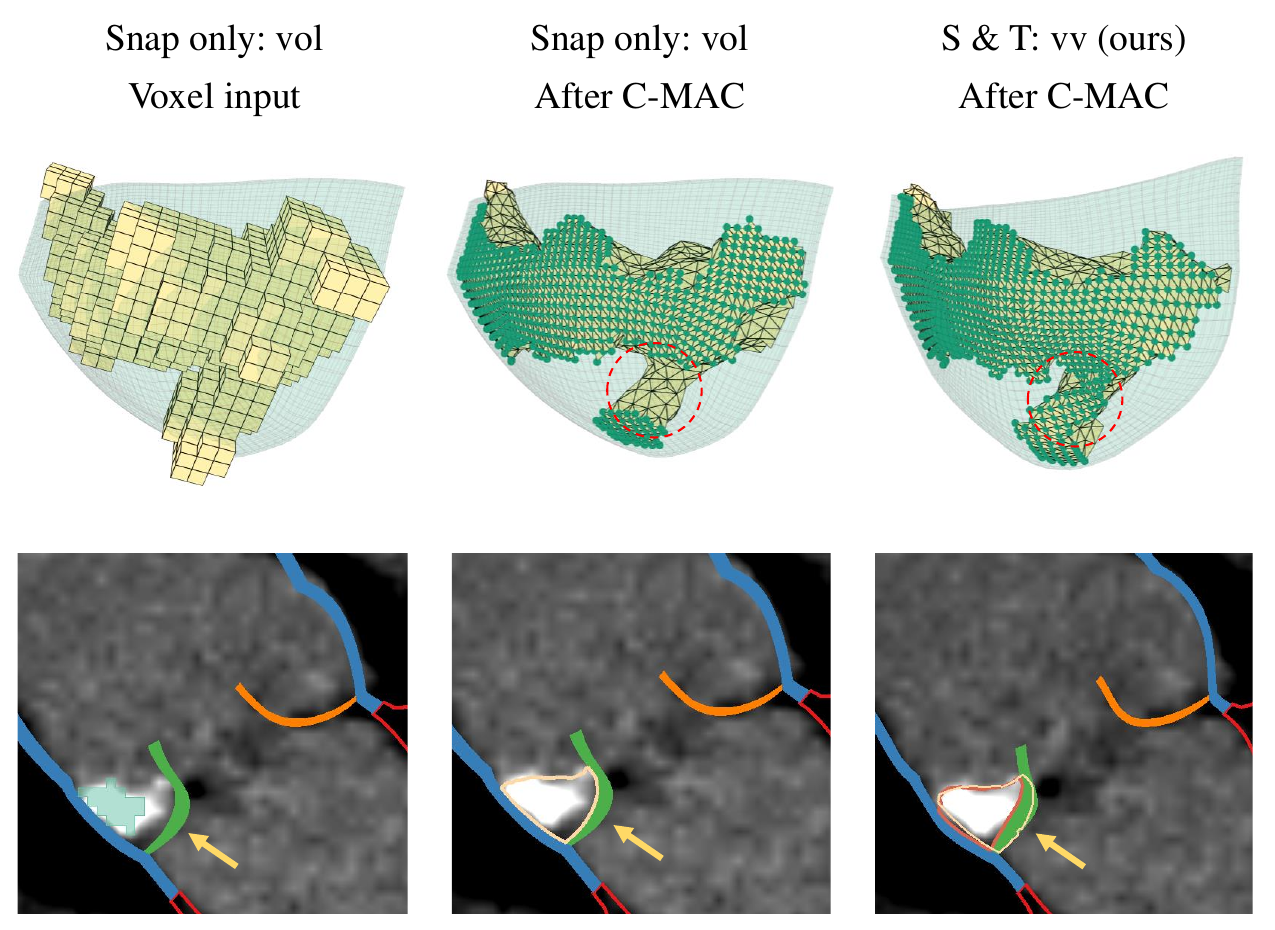}}
\caption{\textbf{Improved mesh attachments.} (Top) Better inter-part consistency indicated by more realistic merged nodes (green dots). (Bottom) Less bleeding indicated by better alignment of segmentation (green voxels) and mesh attachments (yellow and red contours).}
\label{fig:results_ca2_comparisons}
\end{figure}

As shown in Table~\ref{table:ca2}, the overall trend in performance is similar to that of volumetric mesh predictions (Section.~\ref{sec:vol_mesh_pred}). Our method outperforms the previous state-of-the-art by a significant margin, which is also visually apparent in Fig.~\ref{fig:results_ca2_comparisons}. The tune step helps reduce the undesired bridge-like gaps between the mesh attachments and the volumetric mesh, as well as the bleeding effect from attempting to attach to distant volumetric meshes.

Volumetric mesh regularization generally performs better than field smoothness regularization. The only notable difference may be that the ``Tune only: vol" had slightly worse overall performance for mesh attachment evaluations, which further corroborates our argument that our final proposed snap-and-tune strategy is the best performing method amongst our collection of algorithms.

\begin{table}[t!]
\centering

\robustify\bfseries
\def\sym#1{\ifmmode^{#1}\else\(^{#1}\)\fi}

\sisetup{
    detect-weight,
    mode=text,
    separate-uncertainty = true,
    table-align-text-post = false,
    table-align-uncertainty = false,
}

\begin{tabular*}{\linewidth}{
    @{\extracolsep{\fill}}
    l
    S[table-format=1.3(3)]
    S[table-format=4.2(4)] @{}
}

\toprule

{\makecell{Method}} & {\makecell{Dice $\downarrow$}} & {\makecell{\# Merged nodes $\uparrow$}} \\

\midrule

Snap only: smooth & 0.732 \pm 0.035 & 2569.2 \pm 1595.3 \\
Snap only: vol    & 0.738 \pm 0.032 & 2554.3 \pm 1646.0 \\

\midrule[0pt]

Tune only: smooth & 0.760 \pm 0.026 & \bfseries 2694.9 \pm 1638.9 \\
Tune only: vol    & \color{red} 0.761 \pm 0.027 & 2578.7 \pm 1546.1 \\

\midrule[0pt]

S \& T: ss        & 0.751 \pm 0.028 & 2580.3 \pm 1624.7 \\
S \& T: sv        & \color{blue} 0.766 \pm 0.026 & \color{blue} 2667.1 \pm 1672.7 \\
S \& T: vs        & 0.758 \pm 0.024 & 2578.2 \pm 1621.8 \\
S \& T: vv (ours) & \bfseries 0.766 \pm 0.025 $^*$ & \color{red} 2649.7 \pm 1618.6 $^{*\dagger}$ \\

\midrule

\end{tabular*}

\caption{\textbf{Spatial accuracy and anatomical consistency metrics for the mesh attachments.} \# Merged nodes: number of merged nodes between calcification and tissue meshes. Otherwise, same symbols as Table.~\ref{table:heart}.}

\vspace{-0.05in}

\label{table:ca2}

\end{table}

\subsection{Solid mechanics simulations}

We performed large deformation solid mechanics simulations in two different software platforms to demonstrate the versatility and simulation-readiness of our generated meshes.

\begin{figure*}[!t]
\centerline{\includegraphics[width=\linewidth]{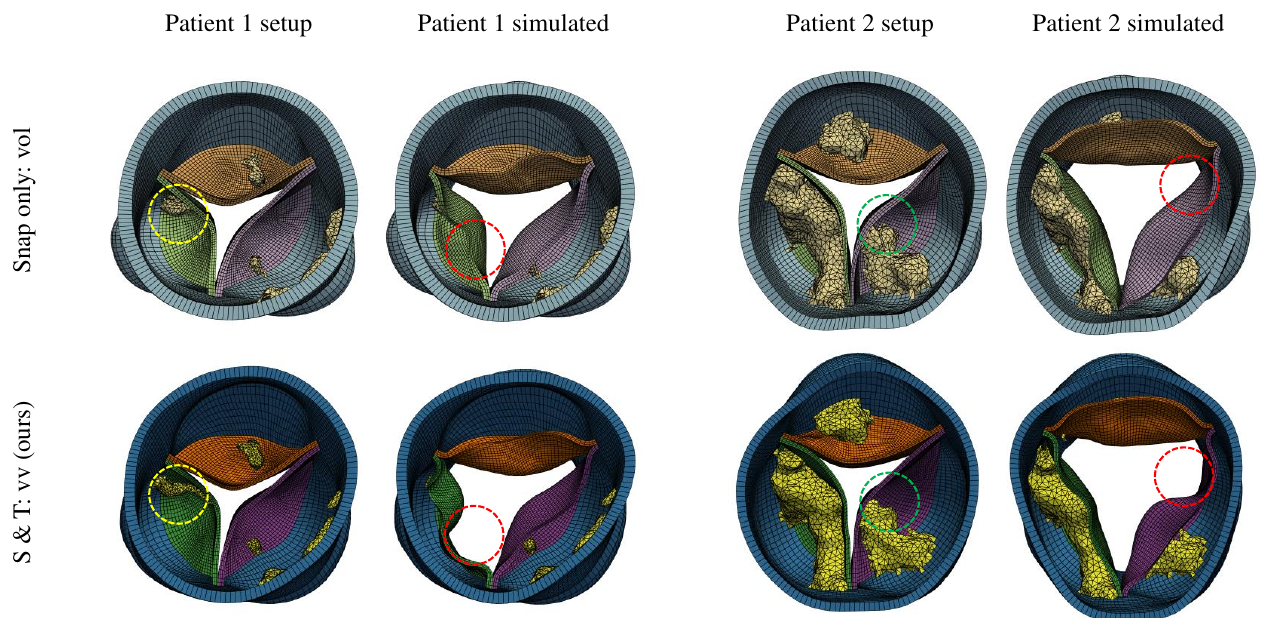}}
\caption{\textbf{High-fidelity engineering simulations (Abaqus).} Relatively small adjustments in the final mesh can have a large influence on the simulation outcome. The reduced bleeding effect of snap-and-tune + C-MAC leads to more realistic valve opening simulations.}
\label{fig:results_simulations_abaqus}
\end{figure*}

First, we performed valve opening simulations on Abaqus (Dassault Systéms), a popular commercial finite element solver used in various high-fidelity engineering applications. Notably, the generated meshes were directly imported into the software without any adjustments. Then, we assigned varying levels of linear elasticity to each mesh component, defined a few fixed boundary conditions to hold the object in place, and applied a uniform normal pressure on the bottom side of the valve leaflets (green, purple and orange meshes in Fig.~\ref{fig:results_simulations_abaqus}). As shown in Fig.~\ref{fig:results_simulations_abaqus}, the seemingly small changes in the mesh attachments (yellow) led to large differences in the simulated valve opening shapes and sizes.

\begin{figure}[!t]
\centerline{\includegraphics[width=\linewidth]{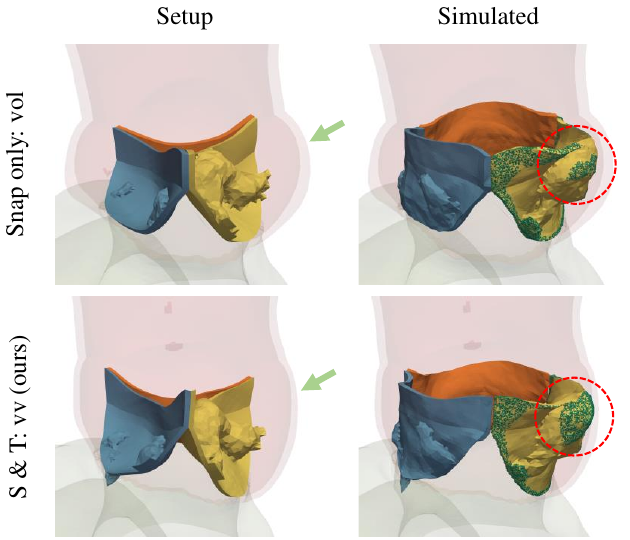}}
\caption{\textbf{Fast graphics simulations (Physx).} Our mesh corrections can heavily influence contact mechanics in device implant simulations. Green dots: contact points.}
\label{fig:results_simulations_physx}
\end{figure}

Then, we performed simplified device implant simulations on NVIDIA PhysX via Omniverse, a platform primarily built for graphics applications that also contains fast physics solvers. The generated volumetric meshes were automatically processed into merged surface meshes for software compatibility. We assigned linear elasticity materials and used the default physics scene parameters, and added a kinematic rigid object to emulate an expanding prosthetic device commonly used in heart interventions. The adjustments in the high-curvature areas led to significant changes in contact mechanics, shown by the distribution of contact points in Fig.~\ref{fig:results_simulations_physx}.

These results show that the snap-and-tune strategy not only improves mesh accuracy from the imaging-based evaluations, but also contributes significantly to the results of the downstream simulations.

\section{Discussion}
\label{sec:discussion}

In this study, we primarily focused on high-fidelity personalized volumetric meshing of cardiovascular structures from patient CT. The improvements were observed over multiple experimental conditions in terms of both spatial accuracy and element quality. These improvements contributed meaningfully to the downstream simulation results, altering the degree of aortic valve opening and the contact mechanics of simplified device deployment. In future studies, we will further assess the effects of the meshing improvements on the simulation accuracy compared to clinical measurements, such as the aortic valve leaflet dynamics vs. echocardiography of aortic stenosis (AS) patients and the post-deployment shape of transcatheter aortic valve replacement (TAVR) devices vs. post-TAVR CT. Furthermore, we will aim to demonstrate the generalizability of our framework with clinical data beyond AS and TAVR patients, first starting with other cardiovascular structures and eventually extending to other imaging modalities and anatomical structures.


From a technical perspective, there is room for improvement in accelerating the test-time optimization step, where different deformation parameterizations and/or optimization techniques may be able to provide performance boosts. Going one step further, it would also be interesting to further investigate the underlying mechanism behind why snap-only predictions tend to be less adaptable to high-curvature areas, and propose corresponding solutions based on network architectural changes or training techniques to address these limitations with DL only. This would obviate the need for time-consuming and potentially user-dependent test-time optimization. A potential solution may involve multi-step DL to reduce the DL methods' tendency to maintain the general structure of the initial template mesh.

Similar to previous explicit template deformation approaches, snap-and-tune has the inherent limitation of not being able to adapt to different topologies of the modeled structure. For AS and TAVR patients, a common congenital defect is bicuspid aortic valves, where we often need to adapt our template mesh to model the patient tissues accurately. Instead of relying on independent template generation for each patient, which would require extensive user input during test-time, we will develop a different but related set of approaches to handle such topological differences automatically using multi-component meshing techniques.
\section{Conclusion}
\label{sec:conclusion}

We proposed a simple yet effective snap-and-tune strategy that provides clear improvements in imaging-based volumetric meshing for personalized cardiovascular simulations. Evaluations were performed over a wide variety of metrics and tasks, including two different solid mechanics simulations of relevant clinical scenarios. In future studies, we will further assess the effects of the mesh improvements on the final simulation accuracy. We will also develop new methods for congenital defects, where  deformation of a single explicit template may not be applicable.



\section*{Declaration of competing interest}
Daniel H. Pak is the inventor of a related provisional patent application No. 63/611,903. There are no other competing interests.

\section*{Acknowledgments}
This work was supported by the National Heart, Lung, and Blood Institute (NHLBI) of the National Institute of Health (NIH), grants T32HL098069 and R01HL121226.

\section*{Data availability}
The data that has been used is confidential.





\bibliographystyle{elsarticle-num} 
\bibliography{refs}







\end{document}